%% file: conference_paper.tex
\documentclass[conference]{IEEEtran}
\IEEEoverridecommandlockouts
\usepackage{cite}
\usepackage{amsmath,amssymb,amsfonts}
\usepackage{algorithmic}
\usepackage{graphicx}
\usepackage{textcomp}
\usepackage{xcolor}
\usepackage{xcolor}
\usepackage{multirow}
\usepackage{booktabs}
\usepackage{url}
\usepackage{hyperref}

\def\BibTeX{{\rm B\kern-.05em{\sc i\kern-.025em b}\kern-.08em
    T\kern-.1667em\lower.7ex\hbox{E}\kern-.125emX}}
\begin{document}

\title{Interpretable Multimodal EEG-ECG Framework for Neonatal Seizure Prediction and Patient-Adaptive Transfer Learning}




\author{\IEEEauthorblockN{Sithmini Ranasingha}
\IEEEauthorblockA{\textit{University of Moratuwa}\\
Moratuwa, Sri Lanka \\
0009-0003-5855-7733}
\and
\IEEEauthorblockN{Agasthi Haputhanthri}
\IEEEauthorblockA{\textit{University of Moratuwa}\\
Moratuwa, Sri Lanka \\
0009-0001-3588-5684}
\and
\IEEEauthorblockN{Hansa Marasinghe}
\IEEEauthorblockA{\textit{University of Moratuwa}\\
Moratuwa, Sri Lanka \\
0009-0006-9738-9260}
\and
\IEEEauthorblockN{Kithmin Wickremasinghe}
\IEEEauthorblockA{\textit{University of British Columbia}\\
Vancouver, Canada \\
0000-0003-1822-649X}
\and
\IEEEauthorblockN{Nima Wickramasinghe}
\IEEEauthorblockA{\textit{University of Melbourne}\\
Melbourne, Australia \\
0000-0001-8133-646X}
\and
\IEEEauthorblockN{Jithangi Wanigasinghe}
\IEEEauthorblockA{\textit{University of Colombo}\\
Colombo, Sri Lanka \\
0000-0002-9413-8363}
\and
\IEEEauthorblockN{Chamira U. S. Edussooriya}
\IEEEauthorblockA{\textit{University of Moratuwa}\\
Moratuwa, Sri Lanka \\
chamira@uom.lk}
\and
\IEEEauthorblockN{Joshua P. Kulasingham}
\IEEEauthorblockA{\textit{University of Moratuwa}\\
Moratuwa, Sri Lanka \\
pranjeevank@uom.lk}
}

\maketitle
\input{FYP/1_Abstract}
\input{FYP/2_Intro_and_Related_work}
\input{FYP/3_Dataset_and_Pre-processing}

\input{FYP/4_Proposed_Model}
\input{FYP/5_Model_Training}

\input{FYP/6_Results_and_Discussion}

\input{FYP/7_Conclusion}

\bibliographystyle{IEEEtran}

\bibliography{FYP/9_References}

\end{document}

%% file: FYP/1_Abstract.tex
\begin{abstract}
Neonatal seizures carry substantial risks of acute brain injury and long-term neurodevelopmental impairment, motivating the need for timely identification of impending seizure activity. While continuous electroencephalogram (EEG) monitoring remains the clinical gold standard, reliable patient-independent seizure prediction in neonates remains largely under-explored. We propose an interpretable multimodal framework that combines EEG and electrocardiogram (ECG) signals for neonatal seizure prediction. Our method uses mel-frequency cepstral coefficient (MFCC) representations processed by a lightweight ($\sim$70k-parameter) convolutional neural network with squeeze-and-excitation attention. The framework was evaluated on the Helsinki neonatal EEG-ECG dataset and the Siena scalp EEG-ECG dataset. Under strict leave-one-patient-out (LOPO) evaluation, patient-independent prediction proved challenging (mean accuracy 58.5\%, F1-score 60.3\%), highlighting substantial inter-subject variability in neonatal EEG. However, fine-tuning the pretrained model with only five minutes of patient-specific data per class improved performance to 95.87\% accuracy and 95.85\% F1-score, demonstrating the feasibility of rapid patient-adaptive transfer learning. SHapley Additive exPlanations (SHAP) identified informative EEG channels that showed partial spatial agreement with clinically reported seizure origins. These findings suggest that fully patient-independent neonatal seizure prediction remains challenging, but that patient-adaptive fine-tuning with a small amount of data may provide a practical pathway toward personalized neonatal seizure prediction systems.

\end{abstract}


%% file: FYP/2_Intro_and_Related_work.tex
\section{Introduction}\label{sec:introduction}

Newborns, particularly during the first 28 days after birth, are highly susceptible to seizures due to the immaturity of their central nervous system \cite{kim2023skin}. Epidemiological studies estimate that seizures occur in approximately 1 to 5.5 per 1000 live births in high-income countries and 1 to 39.5 per 1000 live births in middle-income and low-income countries \cite{wanigasinghe-2024,Ryan2024}. Early detection and intervention are critical, as neonatal seizures are associated with secondary brain injury and long-term neurodevelopmental impairments \cite{Buajieerguli}.

Detecting or predicting neonatal seizures based only on clinical observation in the neonatal intensive care unit (NICU) is challenging, as seizures in this population often manifest with subtle or ambiguous symptoms and may be mistaken for normal infant behavior \cite{mizrahi1987characterization}. Continuous video electroencephalogram (video-cEEG) monitoring is considered the gold standard for neonatal seizure detection \cite{Ryan2024}. However, real-time interpretation requires continuous expert supervision, making the process resource-intensive in many clinical settings.

\begin{figure}[!t]
\centering
\includegraphics[width=\columnwidth]{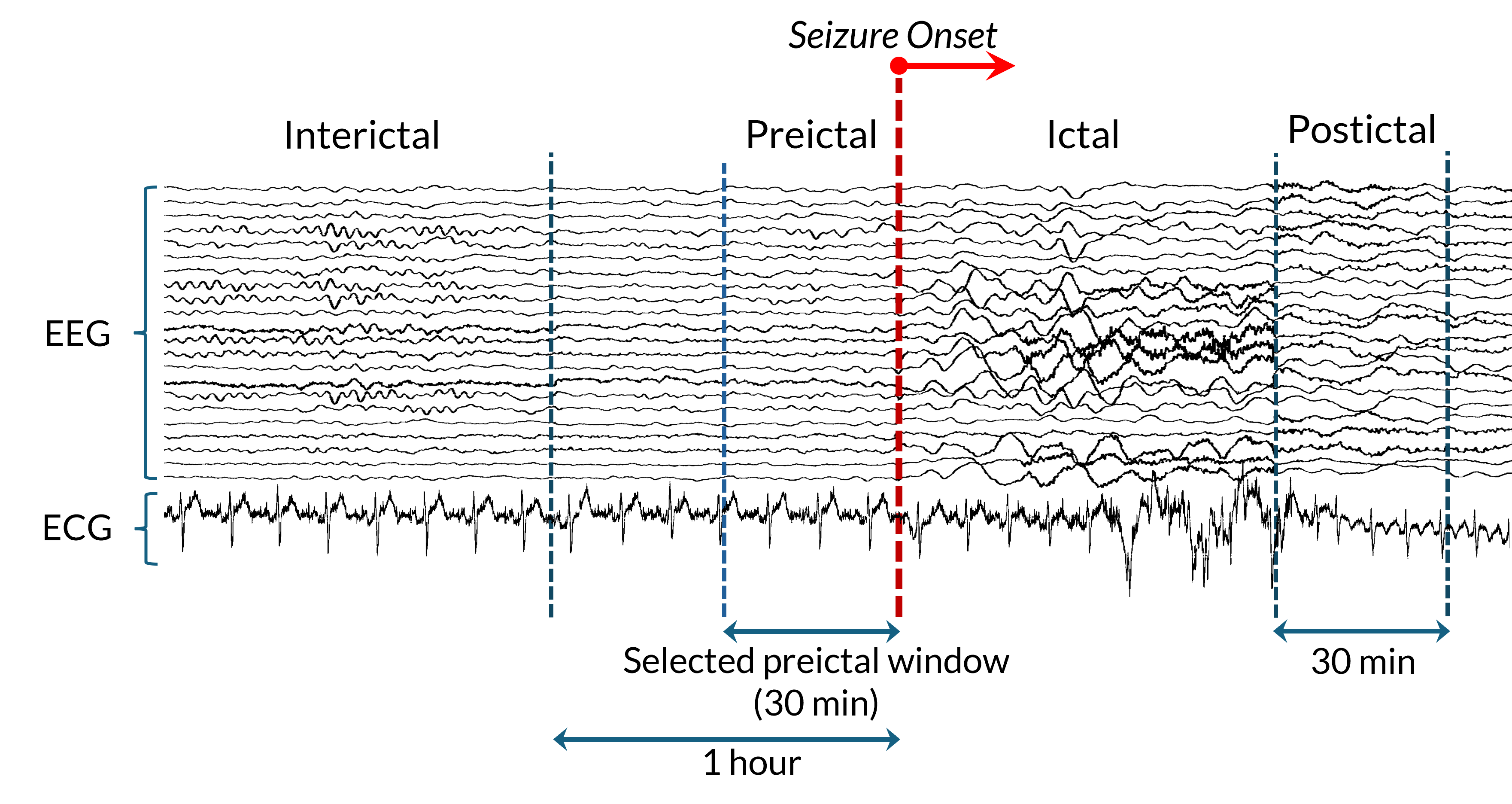}
\caption{\textbf{EEG and ECG recordings across seizure-related brain states.} The figure illustrates interictal, preictal, ictal, and postictal states, along with the time durations considered for the Helsinki dataset.}
\label{seizure_stages}
\end{figure}

EEG signals associated with seizures are commonly categorized into interictal, preictal, ictal, and postictal states, as illustrated in Fig.~\ref{seizure_stages}. While seizure detection focuses on identifying the ictal state, seizure prediction aims to identify the preictal state before seizure onset to enable earlier intervention.

Recent machine learning studies have reported promising seizure prediction performance in pediatric and adult populations \cite{21,theekshana2}. However, neonatal seizure prediction remains relatively underexplored, as most neonatal studies focus on seizure detection rather than prediction. This is challenging because neonatal EEG patterns vary substantially across subjects and because sufficient patient-specific recordings are rarely available.

Seizure prediction approaches are usually categorized as patient-specific or patient-independent. Patient-specific models often achieve high accuracy for a target patient, but they are impractical in neonatal settings due to limited per-patient data. Patient-independent models are more suitable for clinical deployment, but their performance is limited by inter-subject variability in EEG patterns \cite{5,inter_subject_var}. Recent patient-independent seizure prediction studies on pediatric or adult populations have explored convolutional neural networks, recurrent networks, graph neural networks, transformer-based models, and MFCC-based representations for seizure prediction \cite{5,zhang2024mfcc,theekshana2}. MFCCs have shown promise because they provide compact time-frequency representations that can capture seizure-related signal changes.

In addition to EEG, several studies indicate that electrocardiogram (ECG)-derived features can provide useful information for seizure prediction through seizure-related autonomic changes \cite{Mason2024}. Multimodal EEG-ECG analysis has shown promise for neonatal seizure detection \cite{Malarvili2008} and has recently been explored for seizure prediction in adult patients \cite{21}. These findings suggest that ECG may provide complementary information to EEG, particularly when seizure-related EEG changes are not yet strongly expressed. Furthermore, explainability methods such as SHAP are important for improving clinical interpretability and identifying physiologically meaningful model decisions \cite{5}.

To address these gaps, we investigate neonatal seizure prediction using a multimodal EEG-ECG framework that combines patient-independent learning, explainable artificial intelligence, and subject-specific adaptation. The proposed framework employs mel-frequency cepstral coefficient (MFCC) representations extracted from reduced-montage EEG and ECG signals and processes them using a lightweight convolutional neural network with squeeze-and-excitation attention. SHapley Additive exPlanations (SHAP) are used to identify EEG channels contributing to preictal predictions. In addition, we evaluate a transfer learning strategy that enables rapid adaptation to previously unseen neonates using only a small amount of subject-specific data.

The main contributions of this work are summarized as follows:
\begin{itemize}
\item We develop a lightweight multimodal EEG-ECG framework for neonatal seizure prediction and evaluate the contribution of ECG across different prediction window durations.
\item We assess the model under strict leave-one-patient-out evaluation and demonstrate the benefit of patient-adaptive fine-tuning using limited subject-specific data.
\item We incorporate SHAP-based channel-level interpretation to visualize EEG channels contributing to preictal predictions and compare them with clinically reported seizure locations.
\end{itemize}

Overall, this work investigates neonatal seizure prediction beyond detection by combining multimodal physiological monitoring, explainable AI, and patient-adaptive learning, and highlights both the difficulty of subject-independent prediction and the potential value of personalization.

%% file: FYP/3_Dataset_and_Pre-processing.tex
\section{Data Preparation} \label{sec:dataset}

\subsection{Datasets}


The primary dataset used in this study is the Helsinki neonatal EEG-ECG dataset collected at Helsinki University Hospital \cite{zenodo}. Due to the limited availability of neonatal seizure prediction datasets, the proposed framework was additionally evaluated on the Siena scalp EEG-ECG dataset \cite{siena}. The commonly used CHB-MIT dataset \cite{chbmit} was excluded because it lacks ECG recordings and common-reference EEG signals.


The Helsinki dataset contains EEG and ECG recordings from 79 term neonates, including 57 with annotated seizures. To ensure reliable ground-truth labels, only 39 neonates with seizures with consensus annotations from all three clinical experts were retained. Subjects without seizures were excluded because seizure prediction requires learning the physiological differences between preictal and interictal activity within the same subject. Given the substantial inter-subject variability of neonatal EEG, seizure-free recordings provide only interictal data and may bias the model toward learning subject-specific characteristics rather than seizure-related physiological changes. Finally, only neonates with sufficient durations of both preictal and interictal data to construct the prediction dataset were included, resulting in a final cohort of 9 neonates. All recordings were acquired at 256 Hz using the international 10-20 electrode placement system.

The Siena dataset contains EEG and ECG recordings from 14 adults with 47 annotated seizures. Recordings were acquired using the international 10-20 system at 512 Hz and subsequently downsampled by a factor of two.

\begin{figure}
    \vspace{-0.5cm}
    \centering	
    \includegraphics[width=0.2\textwidth]{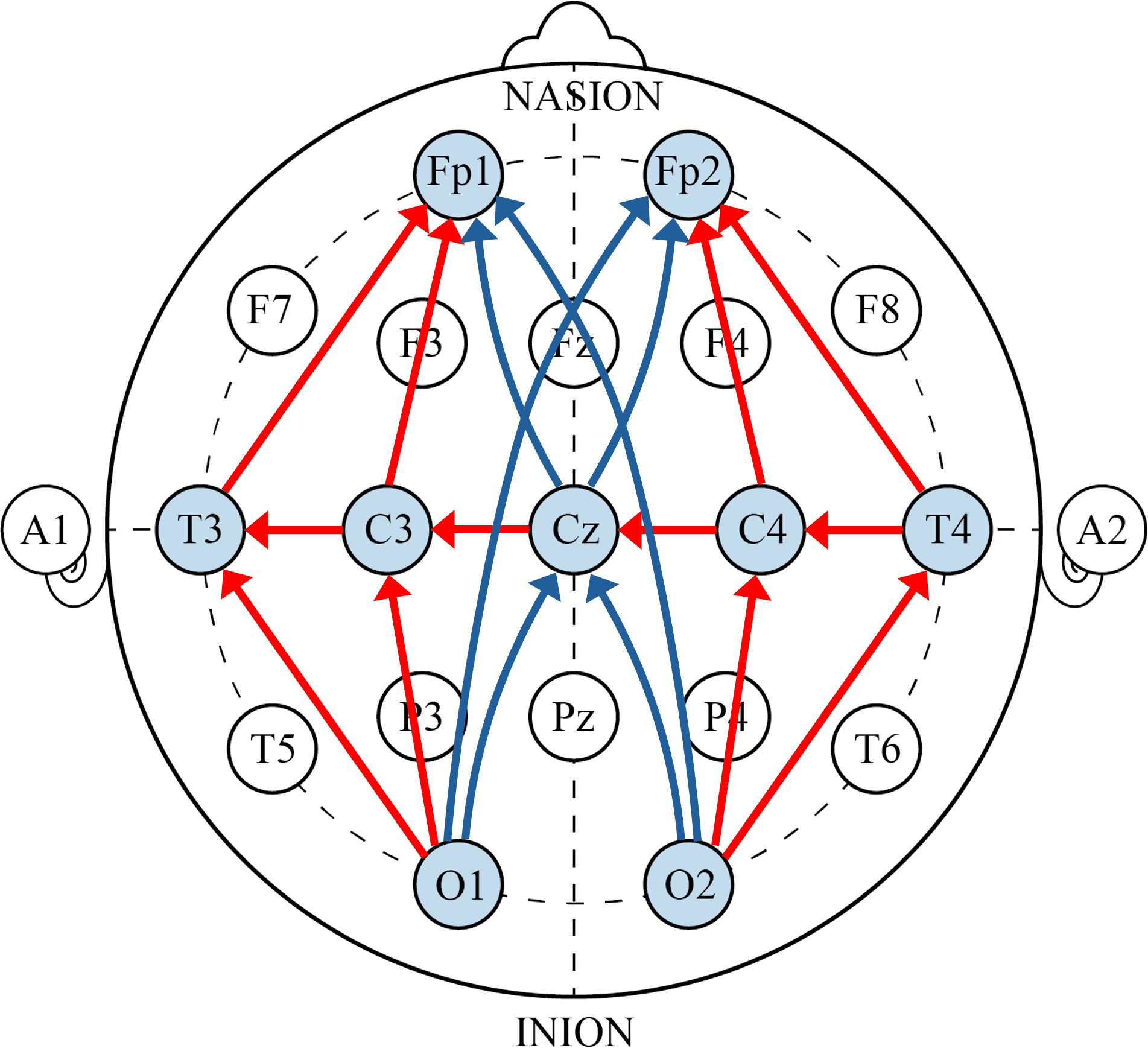}	
    \caption{\textbf{Scalp electrode placement and EEG channel selection.} Selected electrodes are shaded in blue. Red arrows denote the twelve channels from \cite{prev_fyp}, while blue arrows indicate the six additional channels used in this study. Each arrow depicts the electrode pair for each channel, with arrowheads indicating the direction of the difference.}
    \label{electrodes_ML}
\end{figure}

Nine electrodes were selected to construct 18 bipolar EEG channels. Twelve channels were adopted from \cite{prev_fyp}, with six additional channels added to improve spatial coverage. The channel configuration is illustrated in Fig.~\ref{electrodes_ML}.

\subsection{Preprocessing}
\label{sub_sec: preproc}

EEG and ECG signals were filtered using a 4th-order Butterworth bandpass filter (0.1-70 Hz) followed by a 50 Hz notch filter. As no universally accepted preictal duration exists \cite{mid_01}, durations were selected based on prior studies \cite{5,zhang2024mfcc,theekshana2}. For Helsinki, preictal, postictal, and interictal periods were defined as 30 min before seizure onset, 30 min after seizure termination, and at least 1 h before seizure onset, respectively (Fig.~\ref{seizure_stages}). For Siena, a 1 h preictal duration and interictal samples at least 1 h before seizure onset were adopted following \cite{theekshana2}. Recordings were segmented into non-overlapping 5 s windows.

\subsection{MFCC Calculation}

MFCCs have demonstrated effectiveness for EEG-based seizure prediction \cite{5, zhang2024mfcc, theekshana2}. MFCC features were computed from EEG and ECG signals using the Torchaudio library \cite{torchaudio_1}. A hop length of 128 was used, yielding 11 time frames for each 5-second segment, while 20 mel filter banks spanning the 0.5-100 Hz frequency range were employed. Following the data-level fusion strategy in \cite{21}, EEG and ECG channels were stacked along the channel dimension, resulting in MFCC tensors of size $(19,20,11)$ that served as inputs to the proposed model.

%% file: FYP/4_Proposed_Model.tex
\section{Proposed Model Architecture}


Preliminary experiments were conducted using multiple input representations (raw EEG/ECG signals, MFCCs, and statistical features) and deep learning architectures, including CNN-, LSTM-, and encoder-based models. Based on these experiments, the proposed CNN with MFCC inputs was selected because it achieved the best prediction performance while maintaining a compact model. Convolutions over the MFCC representations enable the network to jointly learn spectro-temporal patterns from the EEG and ECG signals while capturing inter-channel relationships across both modalities. Multi-scale convolutional kernels are employed to capture both fine-grained local features and broader contextual patterns. The proposed framework comprises three components: (i) a multi-scale CNN encoder for feature extraction, (ii) a squeeze-and-excitation (SE) attention module for channel-wise feature recalibration, and (iii) a fully connected classifier. The complete architecture, illustrated in Fig.~3, contains approximately 70k trainable parameters. A comparison with existing seizure prediction models is presented in Table~\ref{model_parameters}.

\subsection{CNN Encoder} \label{sec:CNN}

The CNN encoder consists of three sequential 2D convolutional blocks with 32, 64, and 128 filters, respectively. The depth was limited to three blocks because the spatial dimensions of the input MFCC matrices are exhausted after three successive convolution and pooling stages. The channel size of convolution layer in each block was treated as a hyperparameter and tuned by random search, jointly with the remaining hyperparameters, under the 10-fold cross-validation protocol. Each convolution block comprises a convolutional layer followed by batch normalization, ReLU activation, dropout, and max pooling.

To capture seizure-related patterns at different temporal and spectral scales, the first convolutional block employs parallel convolutional kernels with different receptive fields. The resulting feature maps are concatenated and forwarded to the subsequent layers. Multi-scale convolutions are applied only in the first block, while the remaining blocks use a single convolutional branch to maintain computational efficiency.

Batch normalization is used to stabilize training, whereas dropout reduces overfitting. Max pooling with a kernel size of $(2,2)$ progressively reduces the spatial dimensions of the feature maps while retaining salient information. The output of the final convolutional block is subsequently passed to the squeeze-and-excitation attention module.

\subsection{Squeeze-and-Excitation Channel Attention} \label{sec:attention}

Although convolutional layers can automatically learn discriminative representations, not all learned feature channels contribute equally to seizure prediction. This is particularly relevant in multimodal EEG-ECG analysis, where the relative importance of extracted features may vary substantially across subjects and seizure events. To address this challenge, a squeeze-and-excitation (SE) attention mechanism\cite{attention} is incorporated to adaptively prioritize informative channels.

\subsection{Classifier} \label{sec:classification}

The recalibrated feature maps are flattened and passed to a fully connected layer containing 128 neurons. A sigmoid activation function outputs the probability of the preictal class, and a threshold of 0.5 is used for binary classification.

\begin{table}[htbp]
\caption{Comparison of model complexity with representative patient-independent seizure prediction frameworks.}
\centering
\footnotesize
\setlength{\tabcolsep}{10pt}
\renewcommand{\arraystretch}{1.2}
\begin{tabular}{c c}
\hline
{\textbf{Work}} & {\textbf{Approx. No. of Parameters}}\\
\hline\hline 
Dissanayake et al.\cite{5} & 128k\\
Dissanayake et al.\cite{theekshana2} & 289k\\
Zhang et al.\cite{zhang2024mfcc} & 2000k\\
\hline
\textbf{Ours} & 70k\\
\hline
\end{tabular}
\label{model_parameters}
\end{table}

\begin{figure}
    \centering
    \includegraphics[width=0.5\textwidth]{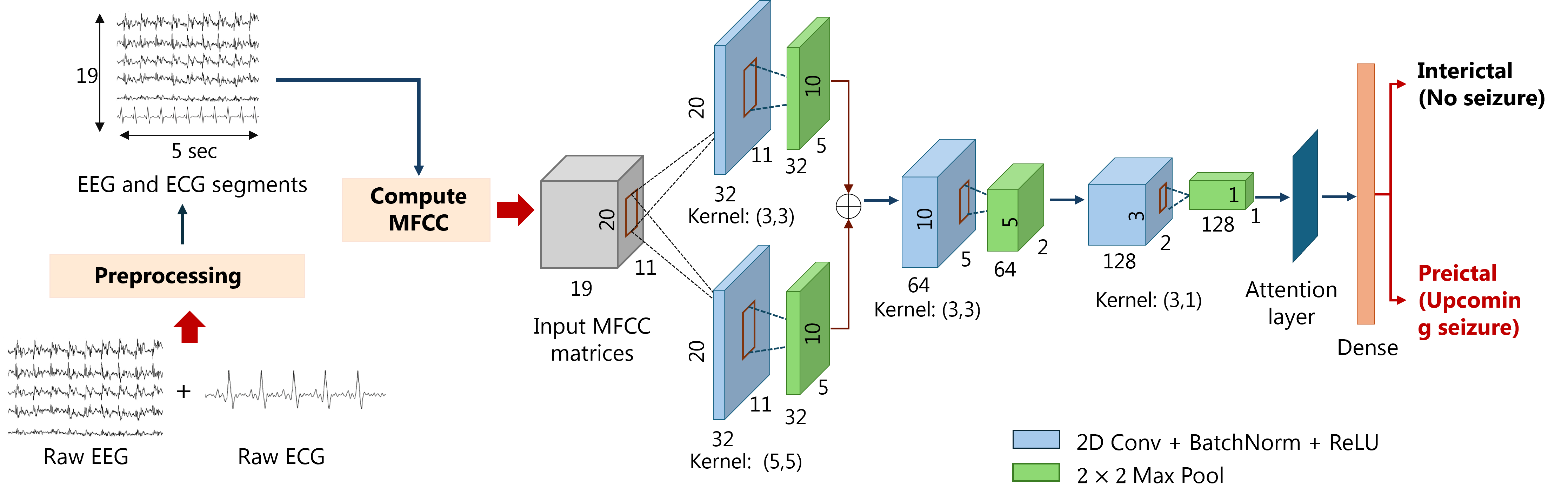}
    \caption{\textbf{Architecture of the proposed deep learning model for seizure prediction.} Raw EEG and ECG are preprocessed, segmented into 5~s windows, and converted into MFCC matrices. Features are extracted using 2D convolutional layers with batch normalization and ReLU (blue) and max pooling layers (green). An attention mechanism precedes the dense classifier, which produces a binary output: 1 for preictal and 0 for interictal.}
    \label{finalmodel}
\end{figure}

The impact of the channel attention mechanism and EEG-ECG fusion strategy is quantitatively evaluated through an ablation study presented in Section~\ref{sec:ablation}.

%% file: FYP/5_Model_Training.tex
\section{Experimental Protocol and Model Interpretation} \label{sec:Training and hyperparameter tuning}

\begin{table}[b!]
\caption{Hyperparameter settings used for model training.}
\centering
\footnotesize
\begin{tabular}{c c c}
\hline
{\textbf{Hyperparameter}} & {\textbf{Helsinki}} & {\textbf{Siena}}\\
\hline\hline 
Learning rate & 0.0004 & 0.002\\
Batch size & 256 & 128\\
Weight decay & 0.005 & 0.0001\\
Dropout & 0.3 & 0.0001\\
Positive class weight & 0.52 & 0.51\\
Reduction factor & 8 & 16\\
\hline

\end{tabular}
\label{hyper_para}
\end{table}

\subsection{Evaluation Protocol}
\label{subsec:eveluation_prtoc}

The primary evaluation protocol in this study was leave-one-patient-out (LOPO) cross-validation, which was used to assess generalization to completely unseen neonates. In each LOPO iteration, all data from one neonate were reserved exclusively for testing, while data from the remaining neonates were used for training and validation. This protocol ensured complete subject separation between training and testing and therefore provided the most clinically relevant estimate of patient-independent neonatal seizure prediction performance.

To investigate rapid adaptation to new patients, the LOPO-pretrained model was subsequently fine-tuned using a small adaptation set from the held-out neonate. Two adaptation scenarios were considered: (i) 2-minute adaptation using 12 preictal and 12 interictal segments, and (ii) 10-minute adaptation using 60 preictal and 60 interictal segments. Adaptation samples were obtained exclusively from the first recorded seizure of the held-out subject and were excluded from subsequent testing. Following fine-tuning, the adapted model was evaluated using the remaining unseen recordings from that subject. Six of the nine neonates satisfied this requirement of having more than one recorded seizure; the remaining three (Neonates 21, 51 and 62) had only a single seizure. For these three subjects, the preictal segments closest to seizure onset were used for adaptation and the most distant segments were reserved for testing, thereby preserving a temporal separation between the adaptation and test segments.  

The model was trained using binary cross-entropy (BCE) loss and the Adam optimizer. Dropout, weight decay, class weighting, and a Reduce-on-Plateau learning rate scheduler were employed to improve generalization and mitigate class imbalance. The weighted BCE loss is defined as,

\begin{equation}
\mathcal{L}_{\mathrm{w}}
=
-\frac{1}{N}
\sum_{i=1}^{N}
\left[
w_{+} y_i \log(\hat{y}_i)
+
(1-y_i)\log(1-\hat{y}_i)
\right]
\end{equation}

where $N$ denotes the number of samples, $y_i \in {0,1}$ is the ground-truth label, $\hat{y}i$ is the predicted probability of the positive class, and $w_{+}$ denotes the weight assigned to the positive class.

Hyperparameters for the 10-fold cross-validation experiments, including the CNN channel sizes, were optimized using random search and fixed prior to evaluation. The final values are summarized in Table~\ref{hyper_para}. For LOPO cross-validation, hyperparameter selection was nested within the outer loop: for each held-out neonate, the hyperparameters were tuned on the remaining eight neonates. The CNN channel sizes were excluded from this nested search and were kept at the values obtained under 10-fold cross-validation, so as to hold the architecture fixed across folds, similar to the hyperparameter tuning procedure in prior work \cite{5}. Model performance was assessed using accuracy, sensitivity, specificity, and F1-score.


Sample-level 10-fold cross-validation was also performed as a secondary experiment to facilitate comparison with existing seizure prediction studies. Five-second EEG-ECG segments generated according to Section~\ref{sub_sec: preproc} were randomly partitioned into ten folds. In each iteration, one fold was used for testing while the remaining samples were divided into training and validation sets using a 70:30 split. Since fold assignment was performed at the sample level, different samples from the same subject could appear in both training and testing partitions. Therefore, the 10-fold cross-validation results should be interpreted as an optimistic estimate of segment-level discriminability rather than as evidence of subject-independent clinical generalization. This limitation is common in the existing patient-independent seizure prediction literature, where sample-level partitioning is frequently adopted.

To assess architecture-level robustness on an additional EEG-ECG seizure prediction dataset, the proposed framework was additionally evaluated on the Siena scalp EEG dataset. The model was trained from scratch using Siena recordings and evaluated using the same sample-level cross-validation protocol with independently optimized hyperparameters. Since the Siena dataset comprises adult subjects, it was used only to validate the model architecture and not to assess its generalizability to neonatal seizure prediction.

\subsection{Model Interpretation}

Interpretability is important in clinical seizure prediction to improve confidence in model decisions and assess their physiological plausibility. To improve interpretability and investigate the spatial distribution of predictive information, SHAP (SHapley Additive exPlanations) \cite{shap} was employed to identify EEG channels contributing most strongly to preictal predictions and to assess correspondence with clinically reported seizure locations.

For each neonate, three preictal segments preceding the earliest seizure with available preictal data and three randomly selected interictal segments were extracted to form an interpretation dataset. A model was trained using the remaining data (70\% training, 30\% validation), and SHAP values were subsequently computed for all interpretation samples. For an input sample of shape $(19,20,11)$, SHAP produces an attribution tensor of identical dimensions. To obtain channel-wise importance estimates, SHAP attribution tensors corresponding to the three preictal segments were averaged for each neonate. Let $\mathbf{S}_{c}\in\mathbb{R}^{20\times11}$ denote the SHAP matrix associated with channel $c$. Channel importance was quantified as
\begin{equation}
I_c = \frac{\mu_{S_c}}{\sigma_{S_c}},
\qquad c \in {1,\dots,19}
\end{equation}

where $\mu_{S_c}$ and $\sigma_{S_c}$ denote the mean and standard deviation of the SHAP values within channel $c$, respectively. Negative importance values were set to zero, the ECG channel was excluded, and min-max normalization was applied to the remaining EEG channel importance values. The resulting normalized importance maps were used to visualize seizure-related spatial patterns.

Our work is publicly available at
\url{https://github.com/Sithminii/BraiNeoCare}.

%% file: FYP/6_Results_and_Discussion.tex
\section{Results and Discussion}\label{sec:Results}

\subsection{LOPO and Patient-Adaptive Transfer Learning Results}

Leave-one-patient-out (LOPO) evaluation was used as the primary and clinically most relevant protocol to assess generalization to completely unseen neonates. In this setting, all samples from one neonate were excluded from training and validation and used exclusively for testing. Under this strict subject-independent setting, the proposed model achieved average accuracy, sensitivity, specificity, and F1-score values of 58.47\%, 58.53\%, 58.14\%, and 60.28\%, respectively, as shown in Table~\ref{table:lopo-results}. Similar subject-independent evaluation by~\cite{5} on the CHB-MIT dataset reported 58.55\% accuracy, which is comparable to our result. These results highlight the difficulty of fully subject-independent neonatal seizure prediction, likely due to substantial inter-subject variability arising from differences in brain maturation, seizure etiology, seizure morphology, and recording conditions.

To address this limitation, we evaluated the patient-adaptive transfer learning strategy described in Section~\ref{subsec:eveluation_prtoc}. The LOPO-pretrained model was fine-tuned using a small amount of labeled data from the held-out neonate and then tested on the remaining unseen data from the same subject. Fine-tuning with one minute of interictal and one minute of preictal data ($N=24$) improved the average accuracy, sensitivity, specificity, and F1-score to 81.55\%, 80.89\%, 84.54\%, and 84.89\%, respectively. Increasing the adaptation data to five minutes per class ($N=120$) further improved these metrics to 95.87\%, 95.25\%, 96.78\%, and 95.85\%, respectively. These findings suggest that, although direct subject-independent prediction remains challenging, the proposed framework can be effectively personalized using limited patient-specific data. This patient-adaptive setting may represent a more feasible approach than fully subject-independent prediction. 

\begin{table*}[h!]
\caption{Subject-wise LOPO and fine-tuned results for each neonate using $N=24$ and $N=120$, corresponding to 1 and 5 min of preictal/interictal data, respectively. Metrics are reported as percentages (\%).}
\centering
\setlength{\tabcolsep}{5pt}
\renewcommand{\arraystretch}{1.1} 
\begin{tabular}{c|cccc|cccc|cccc}
\hline
\multirow{2}{*}{\textbf{Neonate ID}} &  \multicolumn{4}{c|}{\textbf{LOPO}} & \multicolumn{4}{c|}{\textbf{$N=24$ (Int: 1 min, Pre: 1 min)}} & \multicolumn{4}{c}{\textbf{$N=120$ (Int: 5 min, Pre: 5 min)}} \\
\cline{2-13}
 & \textbf{\small{Acc}} &\textbf{\small{Sen}}& \textbf{\small{Spec}} & \textbf{\small{F1-Score}} & \textbf{\small{Acc}} & \textbf{\small{Sen}} & \textbf{\small{Spec}} & \textbf{\small{F1-Score}} & \textbf{\small{Acc}} & \textbf{\small{Sen}} & \textbf{\small{Spec}} & \textbf{\small{F1-Score}} \\
\hline
\hline
11 & 71.92& 65.65& 75.90& 64.37& 95.57 & 93.10 & 97.12 & 94.19 &  95.66& 93.33 & 97.04 & 94.12 \\
13 & 56.83& 36.20& 77.17& 42.83& 83.45 & 86.78 & 80.17 & 83.89 &  91.90 & 91.00 & 92.79 & 91.76\\
15 & 62.09& 62.13& 61.98& 70.39& 70.04 & 64.66& 86.21 & 76.40 & 100 & 100 & 100 & 100 \\
21 & 65.71& 66.25& 64.33& 73.43& 70.98 &75.00 & 60.31& 78.97& 99.48 & 99.67 & 99.33 & 99.74 \\
40 & 46.38& 49.07& 19.45& 57.26& 74.19& 72.70 & 95.83& 84.05& 99.11& 99.04& 100& 99.50\\
51 & 71.84& 76.48& 58.97& 79.93& 83.05 & 82.47 & 84.75 & 87.90 & 91.35 & 92.67 & 85.71 & 94.56 \\
62 & 60.64& 64.90& 58.28& 55.79& 98.94 & 97.00 & 100 & 98.48 & 100 & 100 & 100 & 100 \\
66 & 45.67& 43.98& 69.00& 58.93& 70.42& 70.62& 68.49& 81.23& 89.93 & 89.55 & 100 & 94.48 \\
77 & 45.12& 62.12& 38.17& 39.60& 87.32 & 85.71 & 87.94 & 78.87& 95.37 & 92.00 & 96.17 & 88.45 \\
\hline
Average & 58.47& 58.53& 58.14& 60.28& 81.55 & 80.89 & 84.54 & 84.89& 95.87 & 95.25 & 96.78 & 95.85 \\
\hline
\end{tabular}
\label{table:lopo-results}
\end{table*}

\subsection{Secondary Sample-Level Cross-Validation for Literature Comparison}

\begin{table*}[htbp]

\caption{Sample-level 10-fold cross-validation results. Values are reported as mean ± std across trials.} The sample-level train-test splits do not ensure subject independence. These results are shown in order to compare with prior work.
\centering
\footnotesize
\setlength{\tabcolsep}{6pt}
\renewcommand{\arraystretch}{1.1}
\begin{tabular}{c c c c c c c c}
\hline
{\textbf{Work}} & {\textbf{Dataset}} & {\textbf{Input}} &  {\textbf{Acc (\%)}} & {\textbf{Sen (\%)}} & {\textbf{Spec (\%)}} & {\textbf{F1-score (\%)}} & {\textbf{AUC-ROC}}\\
\hline\hline 

Dissanayake et al.\cite{theekshana2} & Siena & EEG & $96.05\pm 0.17$ & $96.05\pm 0.17$ & $\textbf{96.61} \boldsymbol{\pm} \textbf{0.20}$ & - & 0.99\\

Dissanayake et al.\cite{5} & CHB-MIT & EEG & $91.54\pm 0.17$ & $\textbf{92.45}\boldsymbol{\pm}\textbf{0.22}$ & $\textbf{89.94}\boldsymbol{\pm}\textbf{0.21}$ & -  & 0.96\\

Zhang et al.\cite{zhang2024mfcc} & CHB-MIT & EEG & \textbf{96}& 92 & 84 & 85 & -\\
\hline

\multirow{4}{*}{\textbf{Ours}} & Helsinki & EEG & $96.23\pm 0.20$ & $96.65\pm 0.18$ & $95.58\pm 0.29$ & $96.86\pm 0.17$ & \textcolor{red}{-}\\
& (30 min) & {EEG + ECG} & $\textbf{97.52} \boldsymbol{\pm} \textbf{0.04}$ & $\textbf{98.31}\boldsymbol{\pm} \textbf{0.21}$ & $\textbf{96.39}\boldsymbol{\pm} \textbf{0.33}$ & $\textbf{97.95}\boldsymbol{\pm} \textbf{0.03}$ & 0.99\\

\cline{2-8}

& \multirow{2}{*}{Siena} & EEG & $95.60 \pm 0.09$ & $96.17\pm 0.15$ & $95.02 \pm 0.06$ & $95.62\pm 0.09$ & \textcolor{red}{-}\\
&  & EEG + ECG & $\textbf{96.57} \boldsymbol{\pm} \textbf{0.05}$ & $\textbf{97.31}\boldsymbol{\pm} \textbf{0.01}$ & $95.84 \pm 0.12$ & $96.60 \pm 0.05$ & 0.99\\

\hline
\end{tabular}
\label{result_comparison}
\end{table*}

In addition, sample-level 10-fold cross-validation was conducted on the Helsinki neonatal and Siena adult datasets to facilitate comparison with prior seizure prediction studies that report results under similar cross-validation protocols. The complete evaluation was repeated across three independent trials, and the average results are reported in Table~\ref{result_comparison}. On the Helsinki dataset, the proposed EEG-ECG framework achieved 97.52\% accuracy, 98.31\% sensitivity, 96.39\% specificity, 97.95\% F1-score, and an average ROC-AUC of 0.99. On the Siena dataset, the model achieved above 95\% across all metrics, with an ROC-AUC of 0.99. Despite containing substantially fewer trainable parameters, our method achieved performance comparable to or better than~\cite{theekshana2} across the evaluated metrics on the Siena dataset.

These sample-level results suggest that the proposed multimodal framework can discriminate between preictal and interictal segments when subject-specific patterns are shared across training and testing folds. However, these results should not be interpreted as evidence of subject-independent clinical generalization, because the 10-fold protocol partitions data at the segment level and may therefore include samples from the same subject in both training and testing sets. Accordingly, these results are intended primarily for comparison with existing literature, since prior work uses similar protocols for train-test splits, and should be interpreted as optimistic relative to fully subject-independent evaluation. Although all experiments used strictly non-overlapping 5-second segments to reduce redundancy, the LOPO and patient-adaptive transfer learning results remain the primary indicators of clinical generalization.

\subsection{ECG Fusion, Attention, and Prediction Window}
\label{sec:ablation}

\begin{table*}[htbp]
    
    \caption{Sample-level 10-fold ablation study on the Helsinki dataset. Values are reported as mean ± std across trials.}
    \centering
    \footnotesize
    \setlength{\tabcolsep}{8pt}
    \renewcommand{\arraystretch}{1}
    \begin{tabular}{c c c c c c c }
    \hline
    \multirow{2}{*}{\textbf{Duration}} & \multicolumn{2}{c}{\multirow{2}{*}{\textbf{Input}}} & \multicolumn{4}{c}{\textbf{Performance Metrics}} \\
    \cline{4-7}
    & \multicolumn{2}{c}{} & \textbf{Acc (\%)} & \textbf{Sen (\%)} & \textbf{Spec (\%)} & \textbf{F1-score (\%)} \\
    \hline
    \hline
    
    \multirow{4}{*}{30 min} & \multirow{2}{*}{No attention} & EEG & $95.25\pm 0.08$ & $96.99\pm 0.28$ & $92.55\pm 0.58$ & $96.10\pm 0.07$\\
    & & EEG + ECG  & $96.96\pm 0.05$ & $97.12\pm 0.20$ & $\textbf{96.74}\boldsymbol{\pm}\textbf{0.15}$ & $97.46\pm 0.05$\\
    \cline{2-7}
    & \multirow{2}{*}{Attention} & EEG & $96.23\pm 0.20$ & $96.65\pm 0.18$ & $95.58\pm 0.29$ & $96.86\pm 0.17$\\
    & & EEG + ECG  & $\textbf{97.52}\boldsymbol{\pm} \textbf{0.04}$ & $\textbf{98.31}\boldsymbol{\pm} \textbf{0.21}$ & $96.39 \pm 0.33$ & $\textbf{97.95}\boldsymbol{\pm} \textbf{0.03}$\\
    \specialrule{1.2pt}{0pt}{0pt}
    
    \multirow{4}{*}{15 min}  & \multirow{2}{*}{No attention} & EEG & $\textbf{97.52} \boldsymbol{\pm} \textbf{0.15}$ & $97.62\pm 0.07$ & $97.45\pm 0.21$ & $\textbf{97.22}\boldsymbol{\pm} \textbf{0.17}$\\
    & & EEG + ECG  & $97.25\pm 0.04$ & $97.51\pm 0.23$ & $97.05\pm0.17$ & $96.93\pm 0.04$\\
    \cline{2-7}
    & \multirow{2}{*}{Attention} & EEG & $96.60\pm 0.38$ & $\textbf{98.46}\boldsymbol{\pm} \textbf{0.33}$ & $95.11\pm 0.96$ & $96.28\pm 0.36$\\
    & & EEG + ECG & $97.41\pm 0.05$ & $97.20\pm 0.20$ & $\textbf{97.59}\boldsymbol{\pm}\textbf{0.10}$ & $97.11\pm 0.06$\\
    \specialrule{1.2pt}{0pt}{0pt}
    \multirow{4}{*}{10 min} & \multirow{2}{*}{No attention} & EEG & $97.94\pm 0.11$ & $97.92\pm 0.27$ & $97.97\pm 0.03$ & $97.10\pm 0.15$\\
    & & EEG + ECG  & $97.95\pm 0.50$ & $\textbf{98.68}\boldsymbol{\pm} \textbf{0.24}$ & $97.55\pm0.77$ & $97.18\pm 0.63$\\
    \cline{2-7}
    & \multirow{2}{*}{Attention} & EEG & $\textbf{98.10}\boldsymbol{\pm} \textbf{0.12}$ & $98.24\pm 0.11$ & $98.02\pm 0.13$ & $\textbf{97.32}\boldsymbol{\pm} \textbf{0.16}$\\
    & & EEG + ECG  & $97.99\pm 0.07$ & $97.83\pm 0.24$ & $\textbf{98.08}\boldsymbol{\pm}\textbf{0.16}$ & $97.17\pm 0.11$\\
    \cline{2-7}
    \hline
    \end{tabular}
    \label{ablation}
\end{table*}

To assess the effects of ECG fusion and the attention mechanism, we performed an ablation study using preictal durations of 10, 15, and 30 minutes. For each duration, the model was evaluated with and without ECG input and with and without the attention layer, as summarized in Table~\ref{ablation}.

The benefit of ECG fusion was most evident for the 30-minute prediction window. Without attention, adding ECG increased the F1-score from 96.10\% to 97.46\%, while the full EEG-ECG model with attention achieved the highest F1-score of 97.95\%. For the shorter 15 and 10-minute windows, the performance difference between EEG-only and EEG-ECG models was smaller. This suggests that ECG provides greater complementary value for longer-window prediction, whereas EEG features become more dominant closer to seizure onset. The attention mechanism also improved performance in the 30-minute EEG-ECG setting, indicating that feature recalibration can enhance multimodal prediction when early preictal patterns are more subtle.

\subsection{Model Interpretation}\label{sec:explain}

\begin{figure}
    \centering
    \includegraphics[width=0.85\linewidth, height=7cm]{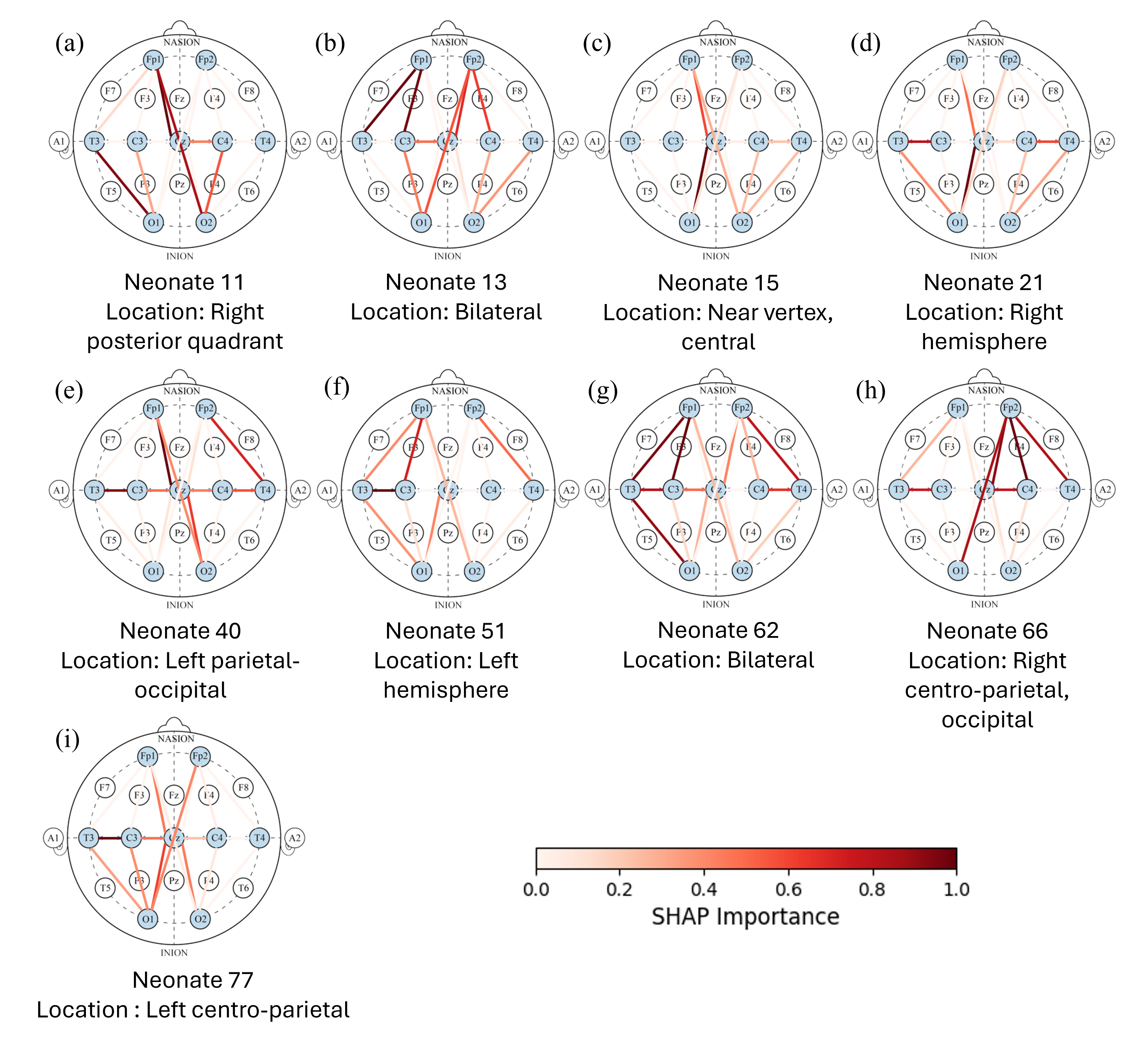}
    \caption{\textbf{Scalp Mapping of SHAP Values for Preictal EEG Samples in Neonates.} The primary seizure location from the dataset metadata is indicated below each plot. Shaded electrodes highlight the nine electrodes used in this study, while lines between electrode pairs represent the 18 EEG channels constructed using the difference of the respective electrodes, as mentioned in Fig. \ref{electrodes_ML}. Line intensity corresponds to SHAP-based importance, with darker lines indicating greater importance.}
    \label{fig:SHAP}
\end{figure}

SHAP-based channel importance values were projected onto scalp maps to visualize the spatial distribution of EEG channels contributing to preictal predictions, as shown in Fig.~\ref{fig:SHAP}. Each line represents one of the 18 bipolar EEG channels, with darker lines indicating higher normalized SHAP importance. The clinically reported seizure locations are shown beneath each scalp map for comparison.

For most neonates, the channels which were assigned higher importance showed partial spatial agreement with the clinically reported seizure regions. In particular, several highlighted channels were located within or near the reported seizure locations, suggesting that the model captured physiologically meaningful preictal patterns rather than relying only on dataset-specific correlations. Some deviations were also observed, which may reflect early propagation of pathological activity or subject-specific variability in neonatal EEG. Overall, the SHAP maps provide an interpretable channel-level visualization of the model’s preictal predictions.

\subsection{Discussion and Limitations}

The results show that fully subject-independent neonatal seizure prediction remains challenging under strict LOPO evaluation, with substantially lower performance than sample-level cross-validation. This gap emphasizes the importance of subject-level evaluation in neonatal seizure prediction and reflects the strong inter-subject variability present in neonatal EEG. However, patient-adaptive transfer learning substantially improved performance using only a few minutes of labeled subject-specific data, suggesting a practical deployment strategy in which a population-pretrained model is personalized after limited patient data becomes available.

The ablation results further suggest that multimodal EEG-ECG fusion is most beneficial for longer prediction windows. ECG provided the largest performance gain for the 30-minute preictal interval, indicating that autonomic changes may contain complementary predictive information before seizure-related EEG patterns become more prominent. As seizure onset approaches, the relative contribution of ECG decreased, suggesting that EEG features become increasingly dominant closer to the ictal period.

Several limitations should be noted. First, the final neonatal cohort included only nine subjects who satisfied the inclusion criteria, and larger datasets are needed for stronger validation. Second, external validation on an independent neonatal seizure prediction dataset was not possible due to the limited availability of public neonatal EEG-ECG datasets with expert seizure annotations. Third, the patient-adaptive transfer learning protocol assumes access to limited labeled subject-specific data, and future work should investigate alternative adaptation strategies. Finally, although SHAP-based scalp maps provided clinically interpretable channel-importance patterns, expert validation of these patterns is needed to assess their clinical utility.

Overall, the findings indicate that neonatal seizure prediction is difficult in a purely subject-independent setting, but that multimodal EEG-ECG modeling combined with patient-adaptive transfer learning provides a promising pathway toward personalized and interpretable seizure prediction in neonatal intensive care.

%% file: FYP/7_Conclusion.tex
\section{Conclusion}
\label{sec:Conclusion}

This study explored the feasibility of multimodal neonatal seizure prediction using EEG and ECG recordings as a proof of concept.
Under strict leave-one-patient-out evaluation, subject-independent performance remained limited, highlighting the difficulty of generalizing neonatal seizure prediction models across patients. However, patient-adaptive fine-tuning using limited labeled subject-specific data substantially improved performance, suggesting that personalization may be a more feasible direction than fully subject-independent prediction. SHAP analysis provided preliminary channel-level visualizations of model predictions, although further expert validation is required. Future work should evaluate the proposed framework on larger neonatal cohorts and improve model adaptation to previously unseen patients.